\documentclass[conference]{IEEEtran}
\IEEEoverridecommandlockouts
\usepackage{cite}
\usepackage{amsmath,amssymb,amsfonts}
\usepackage{algorithmic}
\usepackage{graphicx}
\usepackage{adjustbox}
\usepackage{lscape}
\usepackage{textcomp}
\usepackage{xcolor}
\usepackage{float}
\usepackage{listings}
\usepackage{xcolor}
\usepackage[utf8]{inputenc}
\usepackage{hyperref}

\definecolor{mygray}{rgb}{0.95,0.95,0.95}
\definecolor{keywordcolor}{rgb}{0.0,0.0,0.6}
\definecolor{stringcolor}{rgb}{0.6,0.0,0.0}
\definecolor{commentcolor}{rgb}{0.0,0.5,0.0}

\lstdefinestyle{mypython}{
    backgroundcolor=\color{mygray},
    language=Python,
    basicstyle=\ttfamily\small,
    keywordstyle=\color{keywordcolor}\bfseries,
    stringstyle=\color{stringcolor},
    commentstyle=\color{commentcolor}\itshape,
    showstringspaces=false,
    frame=single,
    tabsize=4,
    breaklines=true
}

\def\BibTeX{{\rm B\kern-.05em{\sc i\kern-.025em b}\kern-.08em
    T\kern-.1667em\lower.7ex\hbox{E}\kern-.125emX}}
\usepackage{graphicx} 
\begin{document}

\title{Secure and Scalable Blockchain Voting: A Comparative Framework and the Role of Large Language Models
\\
}

  \author{
    Kiana Kiashemshaki\textsuperscript{1}, Elvis Nnaemeka Chukwuani\textsuperscript{1}, Mohammad Jalili Torkamani\textsuperscript{2}, Negin Mahmoudi\textsuperscript{3}\\
    \textsuperscript{1}Department of Computer Science, Bowling Green State University, Bowling Green, OH, USA \\
    \textsuperscript{2}School of Computing,  University of Nebraska-Lincoln,  Lincoln, Nebraska, USA \\
    \textsuperscript{3} Department of Civil Environmental and Ocean Engineering, Stevens Institute of Technology, New Jersey, USA \\
       Emails: kkiana@bgsu.edu,
       elvisc@bgsu.edu,
       mJaliliTorkamani2@huskers.unl.edu,
       nmahmoud1@stevens.edu
}

\maketitle

\begin{abstract}
Blockchain technology offers a promising foundation for modernizing E-Voting systems by enhancing transparency, decentralization, and security. Yet, real-world adoption remains limited due to persistent challenges such as scalability constraints, high computational demands, and complex privacy requirements. This paper presents a comparative framework for analyzing blockchain-based E-Voting architectures, consensus mechanisms, and cryptographic protocols. We examine the limitations of prevalent models like Proof of Work, Proof of Stake, and Delegated Proof of Stake, and propose optimization strategies that include hybrid consensus, lightweight cryptography, and decentralized identity management. Additionally, we explore the novel role of Large Language Models (LLMs) in smart contract generation, anomaly detection, and user interaction. Our findings offer a foundation for designing secure, scalable, and intelligent blockchain-based E-Voting systems suitable for national-scale deployment. This work lays the groundwork for building an end-to-end blockchain E-Voting prototype enhanced by LLM-guided smart contract generation and validation, supported by a systematic framework and simulation-based analysis.
\end{abstract}


\section{Introduction}

\IEEEPARstart{T}he digital transformation of governance has increased the demand for secure, transparent, and efficient electoral systems. Voting, as a cornerstone of democracy, faces long-standing challenges such as fraud, limited transparency, logistical inefficiencies, and threats to voter privacy. These challenges are even more significant in large-scale elections, where accurate tallying and secure data handling are essential for public trust~\cite{b1}.

Electronic Voting (E-Voting) has emerged as a potential solution, offering faster vote counting, improved accessibility, and reduced human error. However, current E-Voting systems remain vulnerable to security breaches, privacy violations, and scalability limitations~\cite{b2, b3}. Addressing these concerns is critical to ensure the integrity and reliability of modern elections.

Blockchain technology presents an opportunity to transform E-Voting through its decentralized, tamper-resistant, and transparent nature. By recording each vote as an immutable transaction, blockchain can reduce fraud and increase verifiability~\cite{b4, b5}. Its distributed consensus mechanisms ensure data integrity without relying on a central authority.

Despite these advantages, blockchain-based voting systems face several unresolved challenges. Many rely on energy-intensive consensus models like Proof of Work (PoW), which are unsuitable for national-scale elections~\cite{b6, b7}. Alternatives like Proof of Stake (PoS) and Delegated Proof of Stake (DPoS) improve efficiency but introduce risks such as centralization and collusion~\cite{b8, b3}. Additionally, ensuring voter anonymity while maintaining transparency is technically complex, and existing cryptographic methods often impose high computational costs~\cite{b9, b10}.

Practical barriers, such as interoperability across blockchain platforms and the absence of standardized protocols, further limit real-world deployment~\cite{b11}.

Recent advancements in machine learning have opened up new possibilities for enhancing blockchain-based voting systems. ML has a wide range of applications, including in software engineering \cite{b29}, healthcare \cite{b30, b31}, finance \cite{b32}, production \cite{b33,b34}, cybersecurity\cite{b35}, and now electronic voting. In the context of E-Voting, ML techniques can be used to detect fraudulent behavior, predict potential security breaches, and optimize resource allocation within the network. Additionally, both supervised and unsupervised learning models can support real-time monitoring and anomaly detection, thereby improving the system’s resilience and trustworthiness.

Although prior research has proposed new consensus models, cryptographic enhancements, and decentralized identity systems, existing studies often lack a holistic comparative framework. Furthermore, the potential of Large Language Models (LLMs) to support E-Voting systems through automated contract generation, anomaly detection, and system validation has not been fully explored in this context.

This paper addresses these gaps by presenting a structured comparative framework to assess blockchain-based E-Voting systems across four critical dimensions: scalability, security and privacy, efficiency, and ease of implementation. We analyze the strengths and limitations of various consensus mechanisms, cryptographic techniques, and architectural models to identify performance bottlenecks and recommend practical optimizations.

Our key contributions are as follows:
\begin{itemize}
    \item We present a comparative analysis of blockchain-based E-Voting systems, covering architectural, cryptographic, and consensus aspects;
    \item We propose optimization strategies, including hybrid consensus models, lightweight cryptographic protocols, and decentralized identity frameworks;
    \item We prototype an LLM-guided smart contract development workflow using Remix IDE and Slither to demonstrate the feasibility of integrating AI into blockchain voting systems;
    \item We provide a roadmap for future research, focusing on scalability, LLM-assisted auditing, and pilot deployments.
\end{itemize}

This work offers a foundation for designing secure, scalable, and intelligent blockchain-based E-Voting platforms enhanced by modern AI tools.

\section{Literature Review}

Blockchain-based E-Voting systems have attracted significant attention due to their potential to enhance transparency, integrity, and security in electoral processes. This section categorizes and reviews prior work across five key areas: blockchain architectures, consensus mechanisms, cryptographic solutions, system-level challenges, and real-world deployment barriers.

\subsection{Blockchain Voting Workflow}

\begin{figure}[H]
\centering
\includegraphics[width=0.85\linewidth]{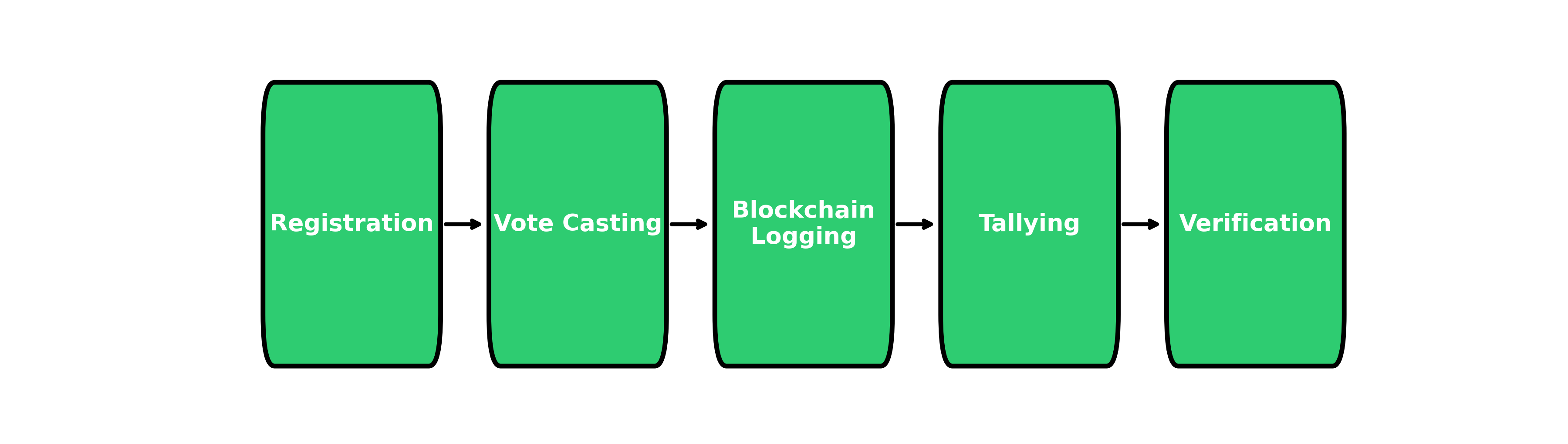}
\caption{Workflow of blockchain-based E-Voting.}
\label{fig:workflow}
\end{figure}

To provide a high-level understanding of how blockchain is integrated into the voting process, we include a visual representation of the main steps in a blockchain-based E-Voting system. Figure~\ref{fig:workflow} illustrates the sequential stages: voter registration, secure vote casting, logging of votes onto the blockchain, tallying, and post-election verification.

\subsection{Blockchain Architectures for E-Voting}

Various blockchain architectures have been proposed to support the secure and scalable implementation of E-Voting systems. Kumar et al. ~\cite{b1} introduced HAC-Bchain, a hybrid system that incorporates dynamic sharding to enhance transaction throughput. While it improves scalability, the model struggles with inter-shard communication, which affects data consistency. Balti et al.~\cite{b4} implemented smart contracts to facilitate transparency in vote recording, but computational overhead hindered scalability when applied to large populations.

Naik et al.~\cite{b2} proposed a permissioned blockchain framework where only verified users can participate. Though this model enhances security and reduces unauthorized access, it risks centralization by relying on a central authority for permission management. Carter and Moore~\cite{b15} emphasized the necessity of establishing standard protocols to enable interoperability between blockchain platforms and electoral systems, a critical factor for national adoption.

\subsection{Consensus Mechanisms}

The consensus algorithm plays a central role in determining a blockchain system’s performance. PoW, while highly secure, is computationally intensive and unsuitable for national-scale voting due to low throughput and high energy consumption~\cite{b3, b6}. Johnson and Patel~\cite{b3} called for transitioning to alternatives like PoS and DPoS, which provide faster block confirmation and reduced energy usage.

However, PoS can lead to disproportionate power concentration among wealthy stakeholders, and DPoS suffers from potential delegate collusion~\cite{b8, b16}. Luu and Wang~\cite{b16} proposed hybrid consensus models that integrate PoS with Byzantine Fault Tolerance (BFT), striking a balance between scalability and security. These models represent a promising path for election systems that require both efficiency and resilience against attacks.

\subsection{Cryptographic Solutions and Voter Privacy}

Ensuring voter anonymity and vote integrity is fundamental in any secure E-Voting system. Singh et al.~\cite{b5} used zero-knowledge proofs and homomorphic encryption to preserve privacy and auditability. Although these solutions are mathematically robust, their computational complexity limits real-time implementation in large-scale elections.

Ahmad and Ahmed~\cite{b17} proposed lightweight cryptographic protocols to reduce resource demands while maintaining confidentiality. Li et al.~\cite{b11} explored decentralized privacy-preserving methods, such as anonymous credentials and self-sovereign identity, to remove the need for central verification authorities while securing voter identity.

\subsection{Challenges and Optimization Strategies}

Blockchain-based voting systems must overcome critical technical barriers, especially those concerning scalability, data throughput, and system efficiency. Bhattacharya and Roy~\cite{b13} identified sharding and parallel processing as scalable approaches, allowing systems to handle large voter volumes by processing multiple chains simultaneously.

Wang and Zhang~\cite{b14} analyzed security vulnerabilities in E-Governance platforms and stressed the importance of resilient protocols against tampering and unauthorized access. Kim and Lee~\cite{b12} performed a comparative evaluation of consensus algorithms, recommending hybrid models for their superior adaptability. Similarly, Yang and Feng~\cite{b18} emphasized the integration of efficient data management techniques into blockchain layers to support the intensive transactional load of election cycles.

\subsection{Real-World Implementations and Future Directions}

Despite growing academic interest, the practical deployment of blockchain-based voting platforms remains limited. Ghobadi and Tavana~\cite{b8} highlighted a lack of standardization and cross-platform compatibility as barriers to adoption. Additionally, national-level systems require collaboration among stakeholders, legal bodies, and technology vendors challenges that current platforms are not fully prepared to meet.

To bridge this gap, researchers advocate for universal standards that ensure platform interoperability, along with regulatory frameworks that promote innovation while safeguarding election integrity~\cite{b7, b15}. Pilot studies and controlled deployments are critical to assess the feasibility of these systems under realistic conditions.

\section{Methodology}

This study employs a structured approach to analyze the current landscape of blockchain-based E-Voting systems. The methodology consists of a systematic literature review, technical performance and security evaluation, the development of a comparative framework, identification of optimization strategies, and recommendations for future research directions.

\subsection{Systematic Literature Review}

A comprehensive literature review was conducted using recent peer-reviewed journals, conference proceedings, and technical reports. The review focused on the following areas:

\begin{itemize}
    \item Blockchain Architectures: Investigating the design and structure of permissioned, permissionless, and hybrid systems with respect to scalability and decentralization.
    \item Consensus Mechanisms: Evaluating the suitability of PoW, PoS, DPoS, and hybrid models for voting use cases.
    \item Cryptographic Protocols: Examining voter privacy techniques such as zero-knowledge proofs, homomorphic encryption, and decentralized identity frameworks.
\end{itemize}

This phase established a foundation for identifying common technical bottlenecks and assessing existing solutions.

\subsection{Technical and Security Evaluation}

The technical analysis was conducted by comparing reported performance metrics, protocol behaviors, and architectural features across a range of E-Voting implementations. Key evaluation areas included:

\begin{itemize}
    \item Consensus Mechanisms: Assessed based on throughput, energy consumption, attack resistance (e.g., Sybil and collusion attacks), and fairness in node participation.
    
    \item Cryptographic Techniques: Evaluated in terms of computational efficiency, scalability, and effectiveness in preserving voter anonymity and vote integrity.
    
    \item Architectural Structures: Analyzed for modularity, fault tolerance, and adaptability to real-world electoral systems. Features such as sharding, decentralized identity, and cross-chain interoperability were considered.
\end{itemize}

\begin{figure}[h]
    \centering
    \includegraphics[width=0.95\linewidth, height=0.35\textheight, keepaspectratio]{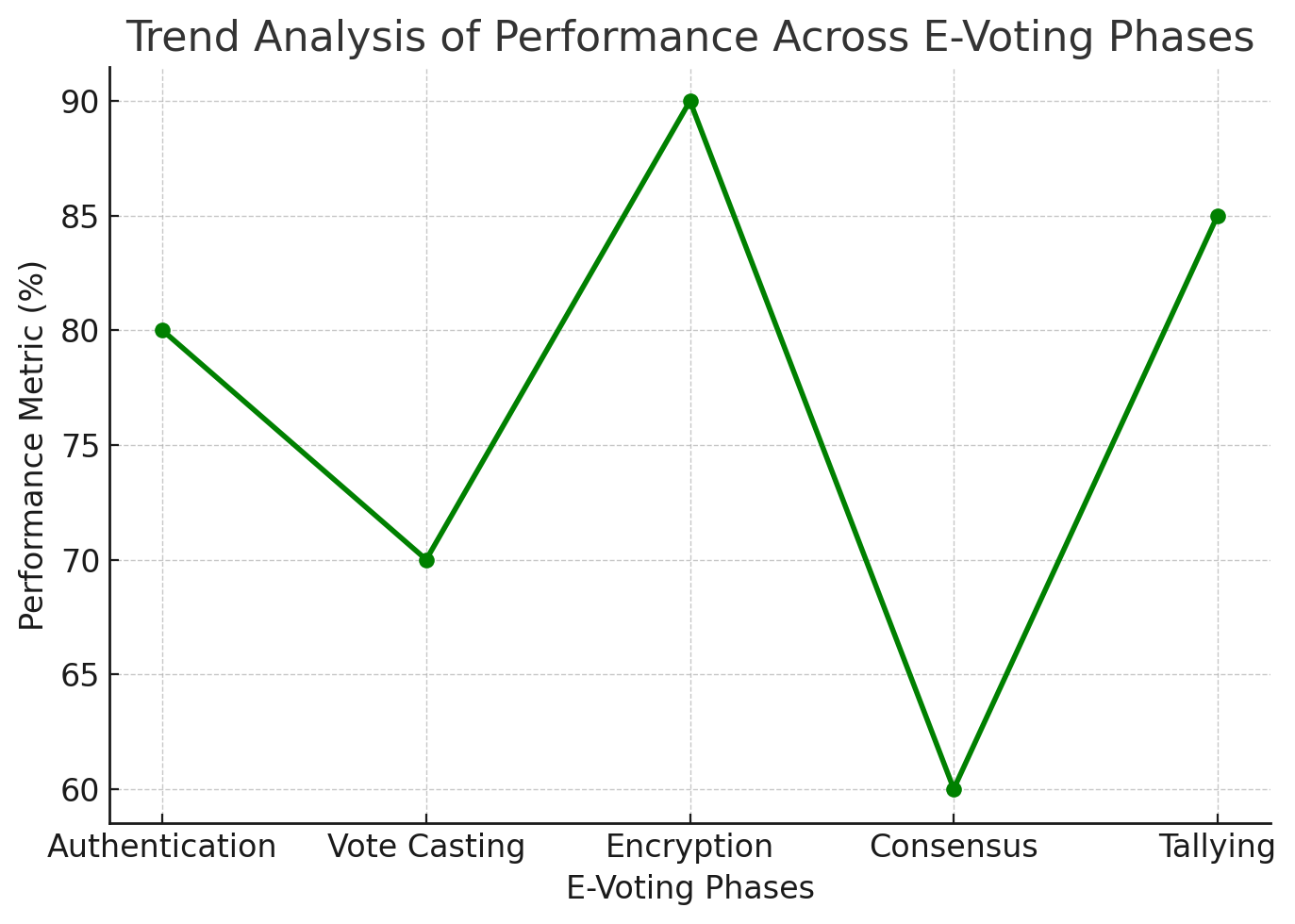}
    \caption{Trend Analysis of Performance Across E-Voting Phases.}
    \label{fig:e_voting_trend}
\end{figure}

\subsection{Comparative Framework Design}

A comparative framework was designed to systematically assess blockchain-based E-Voting systems. The framework includes four core evaluation criteria:

\begin{itemize}
    \item Scalability: The system’s ability to support high transaction volumes typical of national elections. Techniques such as sharding and parallel processing were emphasized.
    
    \item Security and Privacy: Evaluates protections against tampering, data leakage, and identity exposure through the lens of consensus and cryptographic resilience.
    
    \item Efficiency: Measures computational cost, energy consumption, and protocol overhead under various network conditions.
    
    \item Ease of Implementation: Assesses integration complexity, deployment effort, and compatibility with existing electoral systems.
\end{itemize}

This framework enabled objective benchmarking across multiple design configurations and technology stacks.

\subsection{Optimization Strategy Identification}

Based on the results of the comparative analysis, several optimization strategies were proposed:

\begin{itemize}
    \item Hybrid Consensus Models: Combining PoS with Byzantine Fault Tolerance (BFT) or other mechanisms to enhance scalability without undermining decentralization~\cite{b16}.
    
    \item Decentralized Identity Management: Incorporating blockchain-based identity systems to improve privacy, eliminate single points of failure, and streamline voter authentication~\cite{b11}.
    
    \item Lightweight Cryptographic Protocols: Replacing traditional privacy-preserving techniques with computationally efficient alternatives to support large-scale real-time voting~\cite{b17}.
\end{itemize}

\subsection{Future Work and Research Directions}

While this research provides a foundational framework, additional steps are necessary to validate its practical applicability:

\begin{itemize}
    \item Standardization and Interoperability: Future work should support the development of open standards for cross-chain and inter-system compatibility.
    
    \item Machine Learning Integration: AI techniques, including LLMs, can be used to improve system automation, detect voting anomalies, and support smart contract validation.
    
    \item Prototype Development: As a next step, we plan to implement a working prototype featuring hybrid consensus and LLM-assisted smart contract auditing on a private Ethereum testnet.
    
    \item Pilot Studies: Controlled real-world deployments will be necessary to test scalability, user trust, and system resilience under actual voting conditions.
\end{itemize}

\subsection{Prototype Concept: LLM-Assisted Smart Contract Development on a Private Blockchain Testnet}
To demonstrate the feasibility of integrating large language models (LLMs) with blockchain-based voting systems, we designed a prototype that simulates the workflow of LLM-assisted smart contract development and deployment. This proof-of-concept highlights how an LLM can assist in generating a basic Solidity voting contract, which is then tested in a controlled environment.

\subsubsection*{1) Objective}

The goal is to use an LLM, such as OpenAI Codex or GPT-4, to generate a secure voting smart contract that prevents double voting and tracks vote counts. This prototype does not rely on a public blockchain but uses a simulated Ethereum environment to ensure fast deployment and testing.

\subsubsection*{2) Architecture Overview}

The architecture includes:
\begin{itemize}
    \item LLM Interface: A user inputs a prompt like “Write a secure Solidity smart contract for voting.”
    \item Smart Contract Generation: The LLM outputs a Solidity contract with vote tracking and duplicate-vote prevention.
    \item Deployment Environment: The generated code is tested and deployed on the Remix IDE using the built-in VM (e.g., Remix VM Prague).
    \item Execution and Verification: Users interact with the contract by voting, and results are verified on the deployed instance.
\end{itemize}

\subsubsection*{3) Implementation Output}

\begin{figure}[H]
  \centering
  \includegraphics[width=0.9\linewidth]{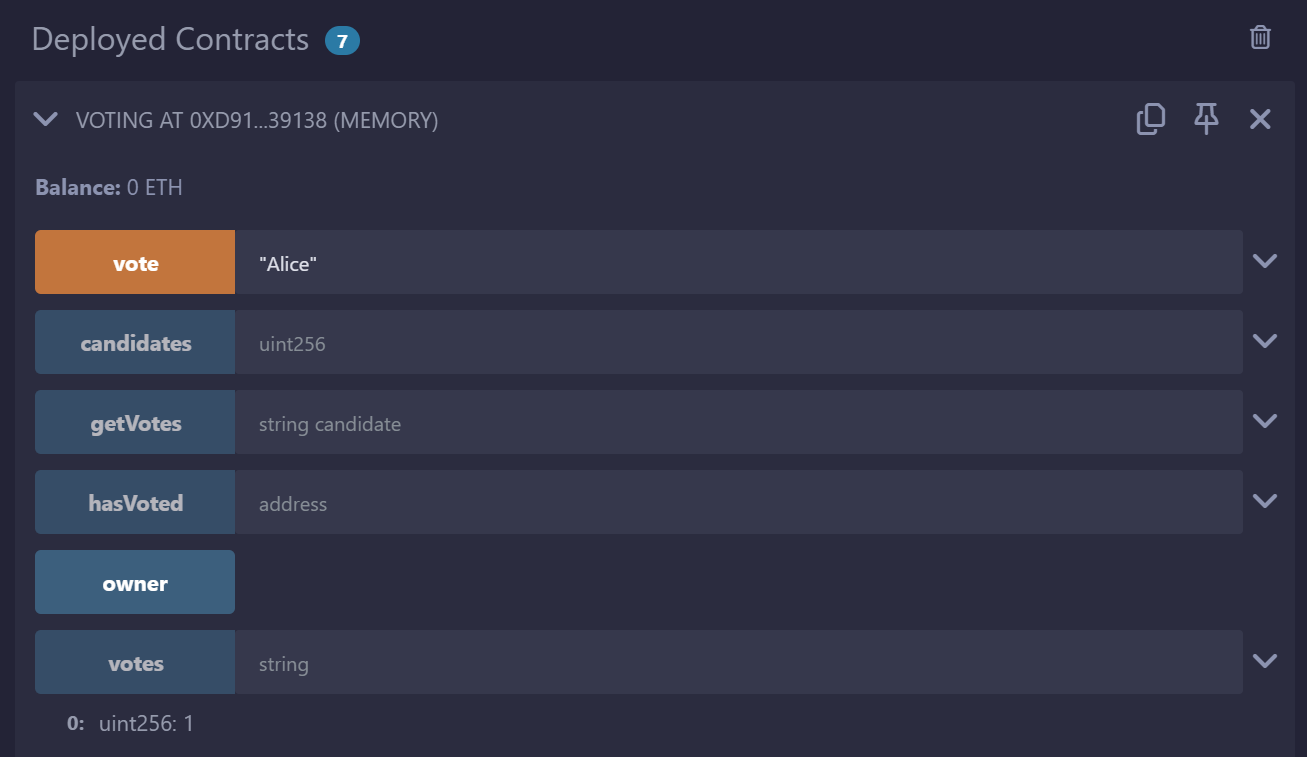}
  \caption{Execution result of the LLM-generated Solidity smart contract on a private Ethereum testnet using Remix. The vote count for “Alice” is incremented after casting a vote.}
  \label{fig:voting-screenshot}
\end{figure}

Figure~\ref{fig:voting-screenshot} shows a successful execution of the smart contract. A vote is cast for the candidate “Alice”, and the contract updates the vote count to reflect the transaction. This confirms that the LLM-generated contract handles voting logic as expected.

\subsubsection*{4) Significance}

This prototype shows that even users with minimal blockchain experience can create and test smart contracts using LLMs. The integration of AI in contract development accelerates experimentation and reduces technical barriers in secure decentralized applications such as E-Voting.

\subsubsection{Objective}

The aim is to automate the transformation of election rules written in natural language into secure, deployable smart contracts. The prototype evaluates each step: from contract generation to security auditing and final deployment on a local blockchain.

\subsubsection{Architecture Overview}

The prototype system includes four core components:

\begin{itemize}
    \item Prompt Interface: Accepts user-defined voting logic in natural language (e.g., one-vote-per-user requirement).
    
    \item LLM Module: Translates the prompt into Solidity smart contract code using LLMs like GPT-4 or Codex.
    
    \item Smart Contract Generation and Testing: The contract is validated using static analysis tools (e.g., Slither) and deployed on a local Ethereum testnet.
    
    \item Blockchain Deployment and Audit: The testnet mimics real-world voting scenarios, allowing for simulation and evaluation of contract behavior under load.
\end{itemize}

\begin{figure}[H]
\centering
\includegraphics[width=0.9\linewidth]{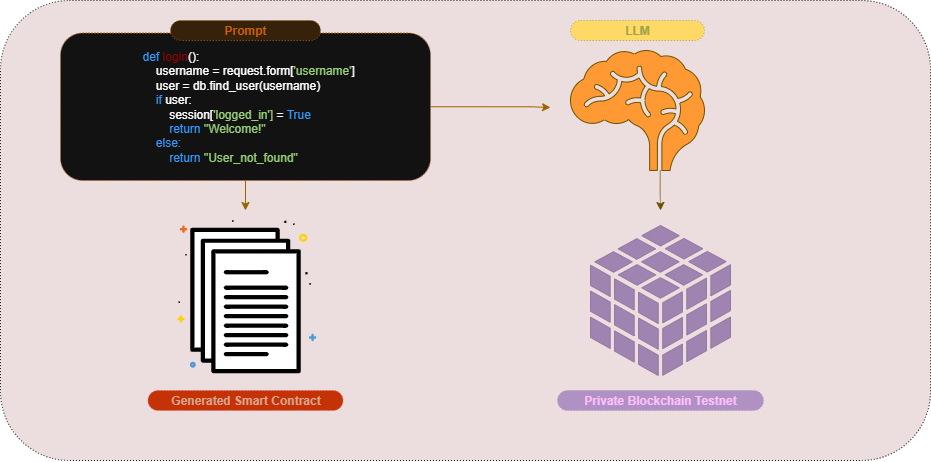}
\caption{System architecture for LLM-guided smart contract generation and testing on a private blockchain testnet.}
\label{fig:llm_testnet_architecture}
\end{figure}

\subsubsection{Advantages and Implications}

This prototype lowers the barrier for secure contract development, allowing non-experts to create and test E-Voting logic with minimal manual coding. By leveraging LLMs alongside testnet deployment and audit tools, the system enables faster iteration, enhanced transparency, and improved stakeholder confidence in blockchain voting protocols.

\subsubsection{Security and Validation}

The LLM output is passed through a static analyzer (e.g., Slither) to flag unsafe patterns, which the system corrects before deployment. Developers are given a plain-language report of audit results.

\subsubsection{Deployment and Testing}

Once validated, the contract is deployed to a local testnet. Voting simulations are conducted using mock voter accounts. Voting behavior is monitored, and logs are analyzed for anomalies using the LLM-auditor agent.

\subsubsection{Benefits and Next Steps}

This prototype reduces developer workload and improves accessibility for non-experts. However, its reliability depends on prompt clarity, model capability, and robust auditing.

Future work includes formal verification of LLM-generated code, integration of zero-knowledge logic for privacy, and real-world pilot deployment under controlled conditions.

\section{Comparative Analysis and Findings}

This section applies the comparative framework developed earlier to analyze existing blockchain-based E-Voting systems. The evaluation spans four dimensions: scalability, security and privacy, efficiency, and ease of implementation. Each dimension is assessed based on current literature and technical designs, leading to key insights that inform future system development.

\subsection{Scalability}

Scalability is a primary challenge for blockchain-based E-Voting, particularly in high-turnout national elections. Systems using Proof of Work (PoW) suffer from low transaction throughput and high latency due to computational bottlenecks~\cite{b3, b6}. These inefficiencies make PoW impractical for time-sensitive voting environments.

More efficient models like Proof of Stake (PoS) and Delegated Proof of Stake (DPoS) offer higher throughput and lower energy consumption~\cite{b8}. However, PoS may centralize voting power among wealthier stakeholders, while DPoS introduces vulnerability to collusion among delegates~\cite{b12, b16}. 

Hybrid consensus models that integrate PoS with Byzantine Fault Tolerance (BFT) mechanisms, such as those proposed by Luu and Wang~\cite{b16}, show strong potential by balancing scalability with decentralization and fault tolerance.

\subsection{Security and Privacy}

While blockchain ensures data immutability, securing voter anonymity and system integrity requires advanced cryptographic techniques. Solutions such as zero-knowledge proofs and homomorphic encryption preserve vote privacy while allowing auditability~\cite{b5, b9}. However, these methods demand substantial computational resources, limiting their usability in real-time elections.

Recent work has introduced lightweight cryptographic protocols to reduce computational overhead without sacrificing privacy guarantees~\cite{b17}. Furthermore, decentralized identity frameworks, as proposed by Li et al.~\cite{b11}, minimize reliance on centralized authorities for voter authentication, reducing potential attack vectors and increasing voter trust.

Systems must also defend against Sybil attacks, double-spending, and node collusion. Consensus models with strong fault tolerance and transparent voting records can help mitigate these threats, but striking a balance between security and system performance remains challenging~\cite{b13, b14}.

Table~\ref{tab:consensus-comparison} provides a high-level comparison of the most commonly used consensus mechanisms in blockchain-based E-Voting systems. Each model is evaluated based on the four key dimensions defined in our framework: scalability, security and privacy, efficiency, and ease of implementation. This comparative overview helps highlight the trade-offs between traditional and emerging approaches, emphasizing the advantages of hybrid consensus models in balancing decentralization, performance, and resilience.

\begin{table*}[ht]
\caption{Comparison of Consensus Mechanisms for E-Voting Systems}
\centering
\begin{tabular}{|l|c|c|c|c|}
\hline
\textbf{Consensus Model} & \textbf{Scalability} & \textbf{Security \& Privacy} & \textbf{Efficiency} & \textbf{Ease of Implementation} \\
\hline
Proof of Work (PoW) & Low & High & Low (Energy-Intensive) & Medium \\
\hline
Proof of Stake (PoS) & Medium & Medium & High & Medium \\
\hline
Delegated PoS (DPoS) & High & Medium (Risk of Collusion) & High & High \\
\hline
Hybrid (PoS + BFT) & High & High & High & Medium \\
\hline
\end{tabular}
\label{tab:consensus-comparison}
\end{table*}

\subsection{Efficiency}

Efficiency encompasses the computational and energy resources required to operate a blockchain voting system. Traditional PoW models, while secure, are highly inefficient due to excessive energy consumption and resource demand~\cite{b6}. This inefficiency poses both scalability and environmental concerns.

In contrast, PoS, DPoS, and hybrid models significantly reduce energy usage and improve confirmation speeds~\cite{b3}. Yet, these improvements must be carefully implemented to avoid compromising decentralization and fairness.

Efficient data management techniques such as sharding, parallel processing, and transaction batching can enhance system performance by enabling concurrent processing of voting data~\cite{b18}. These techniques are critical for scaling systems to handle millions of votes in real time.

\begin{figure}[h]
\centering
\includegraphics[width=0.5\textwidth, keepaspectratio]{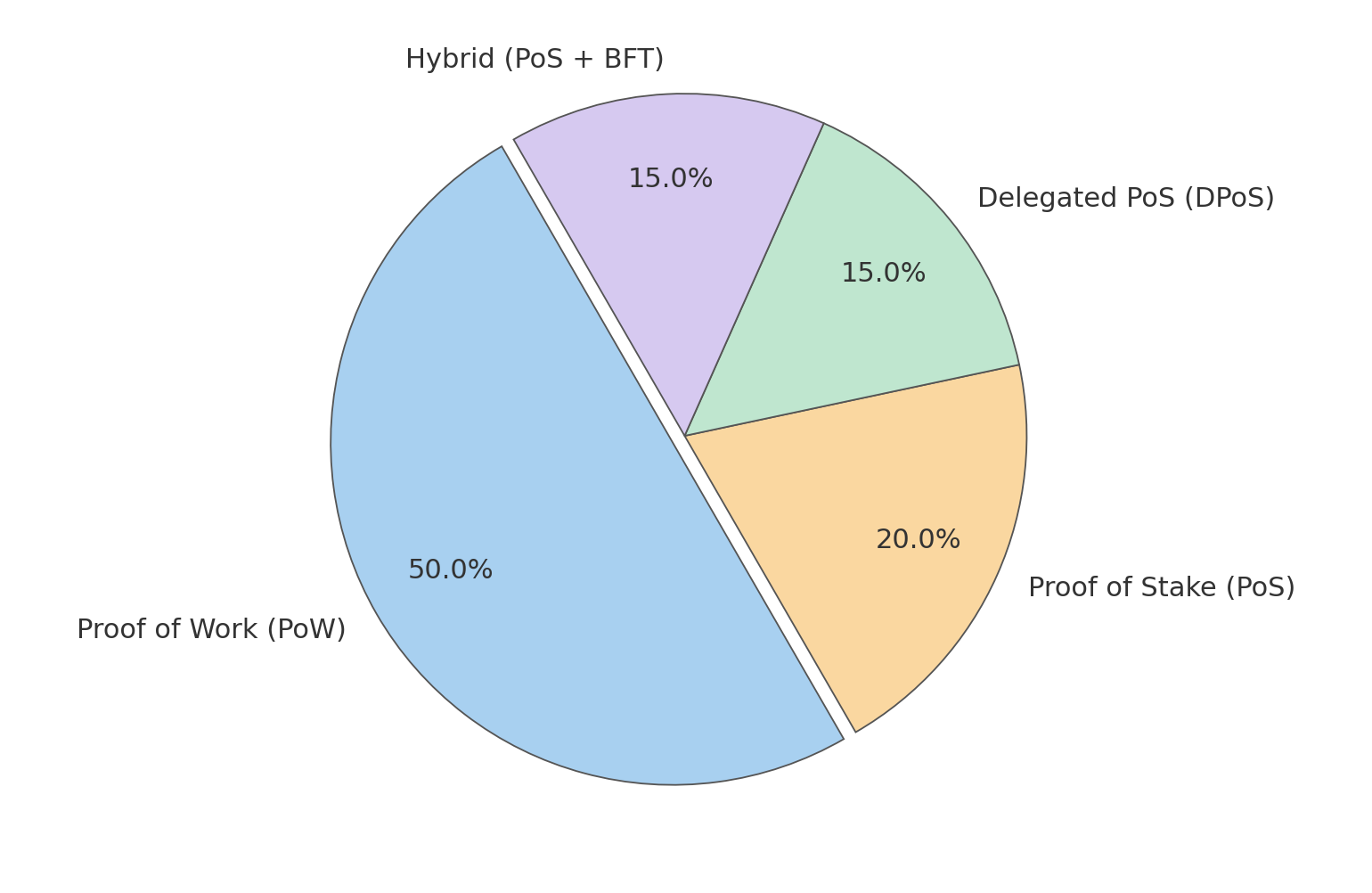}
\caption{Energy Consumption Distribution Among Consensus Mechanisms.}
\label{fig:consensus_pie_chart}
\end{figure}

\subsection{Ease of Implementation}

The practicality of deploying blockchain voting systems depends on both technical complexity and regulatory feasibility. Permissionless blockchains maximize decentralization but are harder to integrate due to infrastructure and compliance challenges~\cite{b1}.

Permissioned blockchains allow faster deployment by restricting participation to verified entities, making them easier to manage. However, this introduces central points of control, which may undermine trust in the system~\cite{b2, b4}. Smart contracts can streamline processes like voter registration and vote tallying~\cite{b4}, though care must be taken to audit them for correctness and fairness.

Emerging technologies like decentralized identity management can enhance implementation by simplifying authentication while maintaining privacy. Standardization across blockchain platforms is also crucial to ensure cross-system compatibility and reduce development overhead~\cite{b11, b15}.

\subsection{Key Findings}

The following insights emerged from our comparative analysis:

\begin{itemize}
    \item Hybrid Consensus is Essential: Integrating PoS with BFT or similar models provides the best trade-off between scalability, energy efficiency, and fault tolerance~\cite{b16}.
    
    \item Cryptography Must Evolve: Lightweight, privacy preserving cryptographic protocols are critical for supporting large scale elections without overwhelming system resources~\cite{b17}.
    
    \item Efficiency Requires Layered Solutions: Combining consensus improvements with advanced data processing (e.g., sharding, parallelism) is key to supporting real-time national elections~\cite{b18}.
    
    \item Implementation Demands Standards: Adoption depends not only on technical design but also on interoperability, smart contract auditability, and regulatory alignment~\cite{b8, b15}.
\end{itemize}

\begin{figure}[h]
\centering
\includegraphics[width=0.7\linewidth, keepaspectratio]{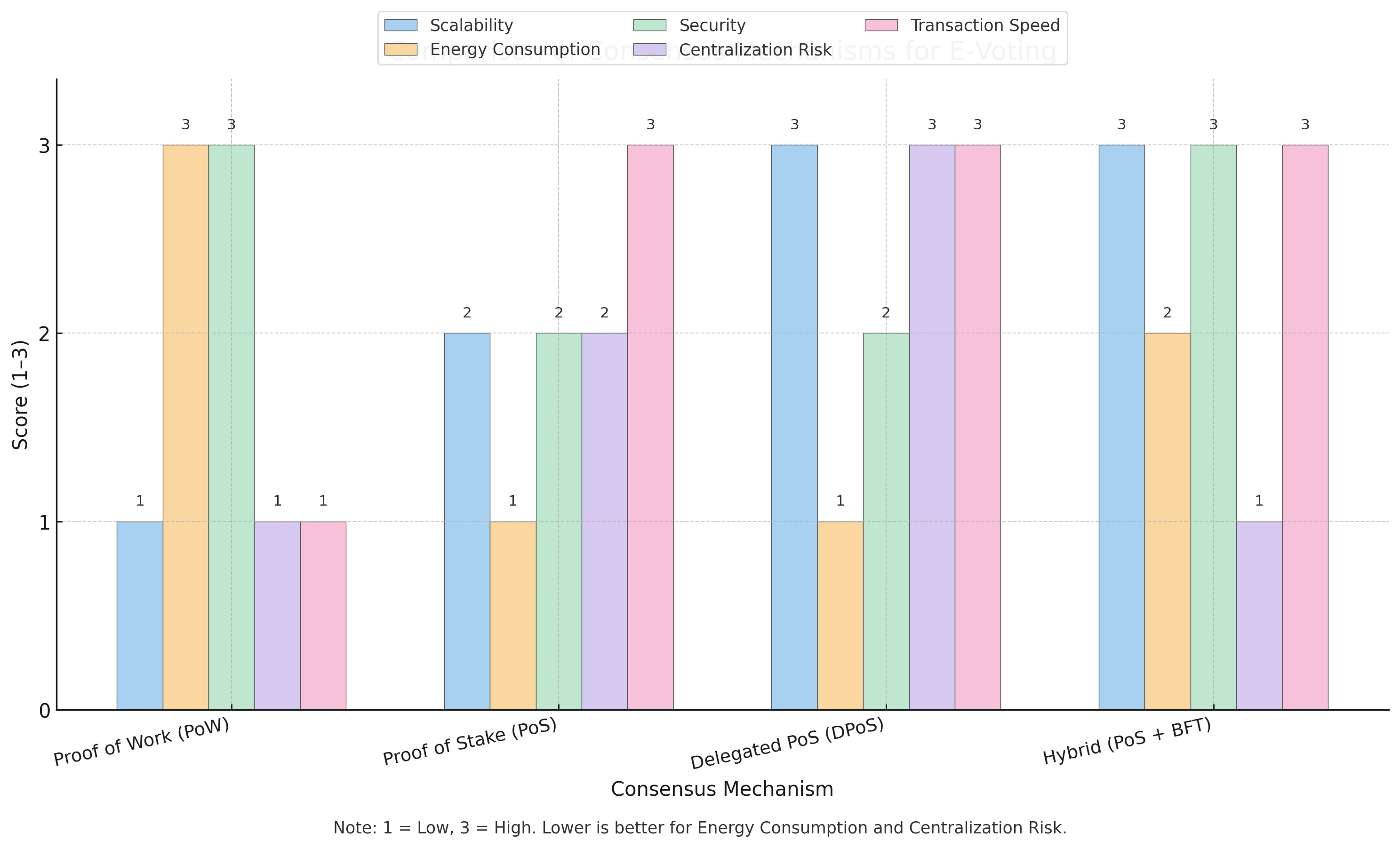}
\caption{Comparison of Consensus Mechanisms for E-Voting.}
\label{fig:consensus_comparison_chart}
\end{figure}

These findings highlight the need for a multi dimensional approach one that integrates secure consensus, efficient computation, and modular architecture to develop future ready blockchain based E-Voting platforms.

\section{The Role of Large Language Models in Blockchain E-Voting}

As blockchain based E-voting systems continue to advance, LLMs are increasingly being recognized for their potential to enhance automation, usability, and overall system robustness. In computer science, LLMs have demonstrated versatility across various domains, including software development \cite{b24, b25, b28}, healthcare \cite{b23, b22}, and education \cite{b26}. Within the realm of software engineering, their capabilities now encompass tasks such as code generation \cite{b27,b21}, documentation, and code review. Furthermore, LLMs have shown promise in identifying and even mitigating software vulnerabilities \cite{b20}. In the context of secure and scalable blockchain systems, this section extends our comparative framework by exploring the role of LLMs in secure smart contract generation, anomaly detection, user interaction support, and system auditability.

\subsection{Enhancing the Comparative Framework with LLMs}

LLMs can contribute to each core dimension of our comparative framework scalability, security and privacy, efficiency, and ease of implementation through various enhancements:

\begin{itemize}
    \item Smart Contract Generation and Optimization: LLMs like OpenAI Codex and GPT-4 can automatically generate secure, optimized smart contracts in languages such as Solidity. This reduces development time, lowers the barrier for non experts, and minimizes manual coding errors~\cite{b19}.

    \item Security and Anomaly Detection: Fine-tuned LLMs can analyze transaction logs and smart contract behaviors to detect anomalies or malicious patterns, offering adaptive fraud detection that outperforms rule based systems.

    \item User Support and Accessibility: ChatGPT style assistants can simplify the user experience by explaining voting steps, system rules, or contract terms in plain language. This supports non technical users and increases system transparency.

    \item Smart Contract Auditing: LLMs trained on secure coding guidelines can review smart contracts to flag potential vulnerabilities or logical inconsistencies prior to deployment~\cite{b19}.
\end{itemize}

\subsection{Emerging Tools and Use Cases}

Recent tools and frameworks demonstrate how LLMs can be integrated into blockchain workflows:

\begin{itemize}
    \item Codex and GPT-4: Used for generating or verifying smart contract code with security annotations.
    \item LangChain and ChainML: Frameworks that bridge LLMs with decentralized applications, enabling dynamic, explainable AI responses within blockchain based systems.
    \item Natural Language to Logic Translation: LLMs can translate voter rules, governance logic, or ballot policies from human language into executable smart contract code, enhancing transparency and reducing legal ambiguity.
\end{itemize}

\subsection{Risks and Limitations}

Despite their advantages, LLMs introduce several challenges when deployed in critical systems such as voting:

\begin{itemize}
    \item Hallucination Risk: LLMs may generate incorrect or insecure code, especially in unfamiliar contexts. Their outputs must be validated using traditional static and dynamic analysis tools.

    \item Bias and Incomplete Training: Models trained on limited or skewed datasets may underperform in fraud detection or contract reasoning, leading to unfair outcomes.

    \item Overreliance on Automation: Excessive dependence on LLM generated code or analysis without human oversight may compromise the integrity and auditability of electoral processes.
\end{itemize}

To mitigate these risks, LLMs should function as decision support tools rather than autonomous agents. Human in the loop frameworks and formal verification methods should accompany any LLM generated outputs.

\subsection{Risks and Mitigation Strategies for LLM Integration}

While LLMs offer powerful enhancements to blockchain based E-Voting systems, they also introduce risks such as hallucination, bias, and overreliance on automation. To address these concerns, we present a matrix of mitigation strategies, evaluated by their effectiveness against each risk.

\begin{figure}[H]
\centering
\includegraphics[width=0.9\linewidth]{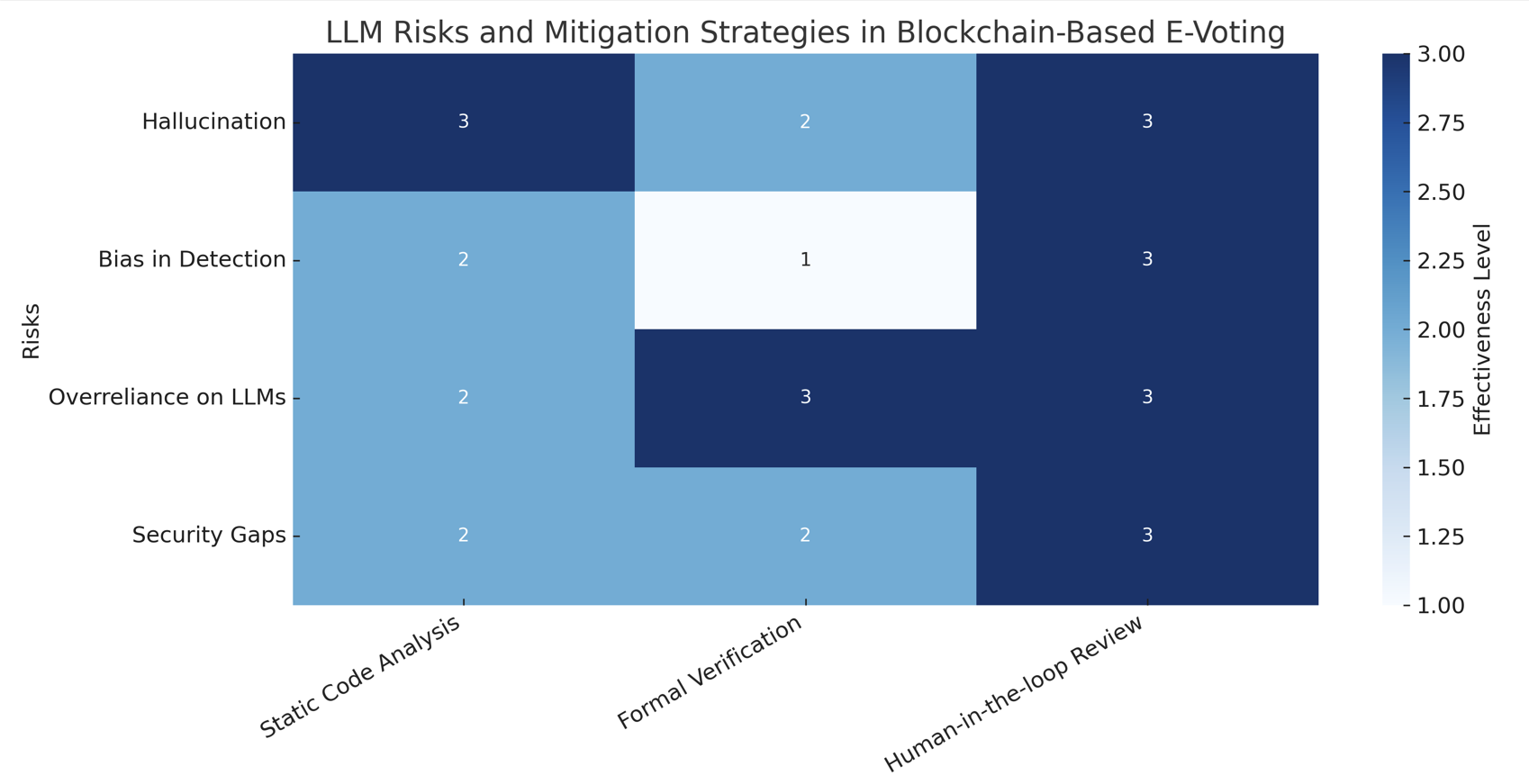}
\caption{LLM Risks and Mitigation Strategies: A heatmap mapping major risks (hallucination, bias, overreliance, and security gaps) to mitigation techniques based on their effectiveness.}
\label{fig:llm_heatmap}
\end{figure}

Figure~\ref{fig:llm_heatmap} illustrates a heatmap of key LLM risks and how targeted mitigation techniques can minimize their impact.

\subsection{Vision for Future Integration}

We envision LLMs serving as modular agents embedded within blockchain voting stacks. Their capabilities can span:

\begin{itemize}
    \item Automated smart contract drafting and policy generation
    \item Natural language explanations of system processes
    \item Continuous security auditing and anomaly flagging
\end{itemize}

\begin{figure}[H]
\centering
\includegraphics[width=0.2\textwidth]{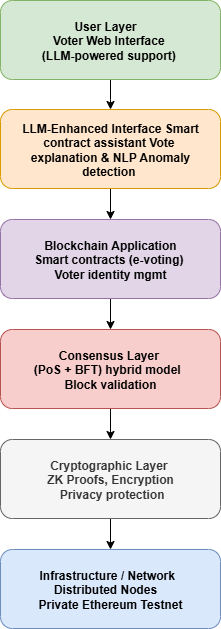}
\caption{System Architecture of a Secure and Scalable Blockchain-Based E-Voting Platform Enhanced with LLMs.}
\label{fig:system_architecture}
\end{figure}

Figure~\ref{fig:system_architecture} presents an architectural view of a secure and scalable E-Voting system enhanced by LLMs, where each layer from user interface to backend logic benefits from intelligent, explainable assistance.

\section{Discussion}

The findings of this study confirm the significant potential of blockchain to modernize E-Voting systems by improving security, transparency, and decentralization. However, practical deployment remains limited by technical, regulatory, and usability challenges. This section discusses the broader implications of our analysis and outlines opportunities for future research and real-world integration.

\subsection{Scalability and Consensus Design}

Scalability is a central concern in national level voting systems. While traditional Proof of Work (PoW) ensures security, its energy consumption and latency make it unsuitable for elections. Hybrid consensus models combining Proof of Stake (PoS) and Byzantine Fault Tolerance (BFT) represent a promising compromise by increasing throughput while maintaining decentralization~\cite{b16}.

However, hybrid models must be carefully designed to prevent new vulnerabilities, such as validator collusion or stake centralization. Future consensus protocols should embed mechanisms that promote fair node distribution, dynamic validator rotation, and adaptive fault tolerance.

\subsection{Security and Voter Privacy}

Ensuring secure vote submission and preserving voter anonymity are critical to system credibility. While blockchain offers data immutability, additional layers such as zero knowledge proofs and homomorphic encryption are necessary to shield voter identities~\cite{b5, b11}. However, these techniques can be computationally expensive.

Lightweight cryptographic protocols~\cite{b17} and decentralized identity solutions~\cite{b11} offer scalable alternatives. By reducing reliance on centralized registrars, these techniques improve voter autonomy and reduce points of failure. Future systems should prioritize cryptographic primitives that enable real-time performance without sacrificing privacy.

\subsection{System Efficiency and Sustainability}

Efficiency extends beyond speed; it includes energy usage, infrastructure costs, and software maintainability. PoS-based systems are notably more sustainable than PoW~\cite{b6}, but require regulatory clarity around token ownership and validator behavior.

Integrating parallel processing techniques, such as sharding~\cite{b18}, can significantly increase throughput while minimizing latency. These techniques should be supported by lightweight virtual machines and modular smart contract environments to ensure long term maintainability and cost effectiveness.

\subsection{Practical Deployment and Standardization}

The transition from research to real-world adoption is hindered by the absence of standardized protocols and legal frameworks. Permissioned blockchains offer easier short term deployment but often compromise on decentralization~\cite{b4, b15}. Interoperability across platforms remains a major barrier.

Collaboration among governments, industry, and academic institutions is necessary to define global standards for E-Voting systems. These standards should cover system interoperability, security validation, and voter authentication. Pilot projects, conducted in controlled environments, are essential to validate system performance under realistic conditions.

\subsection{Limitations and Open Challenges}

While this paper provides a broad comparative analysis, it does not include empirical benchmarking or prototype implementation. Furthermore, the effectiveness of LLMs in voting systems remains largely theoretical and must be evaluated through rigorous testing.

Challenges such as regulatory compliance, hardware requirements in developing regions, and public trust in automated systems remain open for future exploration.

\subsection{Future Directions}

Building upon the results of this study, we recommend the following research directions:

\begin{itemize}
    \item Design and Evaluation of Hybrid Consensus: Develop adaptive hybrid models that dynamically adjust to network conditions while maintaining resilience and fairness~\cite{b16}.
    
    \item Optimized Cryptographic Primitives: Advance lightweight, privacy preserving encryption schemes suitable for real-time voting scenarios~\cite{b17}.
    
    \item LLM Integration and Validation: Build and test LLM powered modules for smart contract generation, fraud detection, and user support in real-world pilots.
    
    \item Interoperability Standards and Governance Frameworks: Collaborate across sectors to define secure, auditable, and legally compliant blockchain voting standards~\cite{b8, b15}.
\end{itemize}

These directions aim to guide both researchers and practitioners toward the realization of secure, scalable, and trusted blockchain based E-Voting infrastructures.

\section{Conclusion}

Blockchain technology holds strong promise for transforming electronic voting systems by offering transparency, security, and decentralization. However, as our analysis has shown, realizing these benefits at a national scale requires overcoming significant technical and regulatory challenges.

This paper presented a comprehensive framework for evaluating blockchain based E-Voting systems. We compared major architectural models, consensus mechanisms, and cryptographic protocols using four key dimensions: scalability, security and privacy, efficiency, and ease of implementation. Our findings underscore the potential of hybrid consensus mechanisms, lightweight cryptographic techniques, and decentralized identity systems in addressing existing limitations.

In addition, we explored the emerging role of Large Language Models (LLMs) in supporting smart contract generation, fraud detection, and voter guidance. These tools can significantly enhance system usability and trust when paired with rigorous validation and human oversight.

To advance this field, we recommend prioritizing the following research directions:
\begin{itemize}
    \item Development of adaptive, secure hybrid consensus models~\cite{b16};
    \item Design of scalable, privacy-preserving cryptographic protocols~\cite{b17};
    \item Standardization efforts to ensure interoperability and legal compliance~\cite{b15};
    \item Integration and testing of LLMs as auxiliary tools in E-Voting pipelines~\cite{b19}.
\end{itemize}

\subsection{Final Thoughts}

The future of E-Voting lies at the intersection of secure blockchain architectures and intelligent automation. By combining decentralized infrastructure with explainable AI tools, governments and developers can create voting systems that are not only transparent and efficient but also adaptable to the evolving demands of digital governance.

Our research provides a solid foundation for such development and encourages further interdisciplinary collaboration to bring secure and scalable blockchain-based E-Voting from concept to reality.

\section*{Acknowledgment}
This paper originated from a group work in the Computer System Security course in the Bowling Green State University. We thank Kyle Cusimano, Uchechi Nwala, and Jacob Partin for their contributions to the original class submission.

\section*{Author Contributions}
K. Kiashemshaki led the concept, framework, prototype, figures, and original draft; E. N. Chukwuani co-led the concept, contributed to methodology, validation, and editing; M. J. Torkamani contributed LLM/security methodology, auditing, validation, and editing; N. Mahmoudi contributed modeling for real-world deployment, validation, and editing. All authors approved the final version.

\section*{Funding}
No dedicated funding was received for this project.

\section*{Conflict of Interest}
The authors declare that they have no conflicts of interest.

\end{document}